\documentclass[reprint,showpacs,preprintnumbers,showkeys,amsmath,amssymb,aps,pra]{revtex4-1}
\usepackage{sankar-book,sankar-draw}
\begin{document}\preprint{APS/123-QED}
\title{Shot noise limited interferometry for measuring classical force}
\author{Sankar Davuluri$^1$}
\email{sd3964@csrc.ac.cn}
\author{Yong Li$^{1,2}$}
\affiliation{$^1$Beijing Computational Science Research Center, Beijing 100193, P. R. China}
\affiliation{$^{2}$Synergetic Innovation Center for Quantum Effects and Applications, Hunan Normal University, Changsha 410081, China.}
\date{\today}
\begin{abstract}We propose an interferometry technique, by using electromagnetically induced transparency phenomena, for measuring classical force. The classical force is estimated by measuring the phase at the output of the interferometer. The proposed measurement mechanism satisfies quantum non-demolition measurement conditions leading to back-action evasion. We further derive the sufficient condition under which the thermal noise in the interferometer is negligible. With no back-action noise and no thermal noise, the sensitivity of this technique is limited by shot noise only.\end{abstract}
\maketitle
\section{Introduction}
Interferometry is a vital component in precision measurement schemes. For example, the gravitational wave detector LIGO is a giant Michelson-Morley interferometer which can detect tiny changes in displacement. While LIGO requires a huge set-up, the recent advances in the field of optomechanics~\cite{aspelmeyer-rmp} can lead to portable ultra-high precision measurement devices. The precision measurements in both the `\textit{huge}' LIGO and the `\textit{tiny}' miniaturized optomechanical cavities are limited by similar fundamental quantum aspects~\cite{clerk,caves}, like shot noise, back-action noise, and thermal noise. Back-action noise limits the measurement precision of the apparatus to standard quantum limit (SQL)~\cite{caves-rmp}. So observation and evasion of back-action noise, along with non-classical states~\cite{yuen} and quantum entangled states~\cite{dowling,ma}, has been studied~\cite{nogues,pontin} extensively in different systems~\cite{milburn,kimble,caniard,qi,moller,lupascu,qixiao,sankar-pra3} in-order to overcome the SQL. Recent advances in experimental physics has lead to observation~\cite{murch,corbitt,spethmann} of back-action noise and standard quantum limit in several systems. In this article, we present interferometry technique, by using three-level atoms~\cite{budker}, for measuring classical force with out back-action noise~\cite{braginsky,braginsky-rmp,grangier,kimble} and thermal noise. The back-action evasion is achieved by using electromagnetically induced transparency (EIT)~\cite{harris,boller,fleischhauer-rmp,novikova} phenomena. We further describe the sufficient condition under which the thermal noise can be suppressed completely by using cold atoms~\cite{hansch}. Because bothe thermal noise and quantum back-action noise are suppressed, shot noise is the only limiting factor in this new interferometer.
\section{Back-action evasion}
\begin{figure*}[htb]
\begin{tikzpicture}[scale=1,spring/.style={|-|,thick,decorate,decoration={coil,aspect=0.8,amplitude=1mm, segment length=1.3mm,post length=1mm, pre length=1mm}}] 
\node[rectangle,draw=black,line width=0.3mm,fill=green!10!,inner sep=2.5mm, minimum size=2.5] (b) at (-4,-1) {LS};
\node[rectangle,draw=black,fill=green!10!,inner sep=2.5, line width=1](c) at (6,2) {D1};
\node[rectangle,draw=black,fill=green!10!,inner sep=2.5, line width=1] (d)at (5,3) {D2};
\draw[line width=1mm,blue!80!](b)--(.48,-1)node[above]at(-1.75,-1){$\hat a_{in}$}node[above,below,gray]at(-1.75,-1){arm-1};
\draw[line width=1mm,blue!80!](5,-1.3)--(d)node[]{};
\draw[line width=1mm,blue!80!](-3,-1)--(-3,2)node[]{};
\draw[line width=1mm,blue!80!](-3,2)--(c)node[above]at(-1.75,2){$\hat a_1$}node[below,gray]at(-1.75,2){arm-2};
\draw[red,fill=red,opacity=0.5](0,-1) circle (0.1)node[]{};
\draw[line width=2, color=blue!50!](-3.2,-1.2)--(-2.8,-0.8)node[]{};
\draw[line width=2, blue!50!](4.8,1.8)--(5.2,2.2)node[]{};
\draw[line width=2](4.8,-1.5)--(5.2,-1.1)node[]{};
\draw[line width=2](-2.8,2.2)--(-3.2,1.8)node[]{};
\draw[line width=2](-0.2,.3)--(0.2,.3)node[above,midway]{2};
\draw[line width=1mm,blue!80!](0.5,-1)--(0,.2)--(-0.6,-1.3)--(5,-1.3)node[]at(-0.35,-0.1){$\hat c$}node[above]at(1.75,-1.3){$\hat a$};
\draw[line width=2, blue!50!](-0.3,-1.2)--(-0.7,-0.8)node[black]at(-0.8,-0.6){m};
\draw[line width=2](0.35,-1.2)--(0.65,-0.8)node[black]at(0.8,-0.6){1};
\filldraw[yellow,fill=yellow](-0.03,-1.03)--(0.03,-1.03)--(0.03,-0.97)--(-0.03,-.97)--cycle node[]{};
\filldraw[yellow,fill=yellow!10!](4.4,0.3)arc(0:360:1.1);
\draw[thick] (3.5,-0.2) -- (4,-0.2) node[below,midway] {$|b\ket$};
\draw[thick] (3,1) -- (3.5,1) node[right] {$|a\ket$};
\draw[thick] (2.5,0.2) -- (3,0.2) node[midway,below] {$|c\ket$};
\draw[thin,blue,<->,>=stealth] (3.75,-0.2) -- (3.27,1) node[midway,right] {$\hat c$};
\draw[thick,red,<->,>=stealth] (2.75,0.2) -- (3.2,1) node[midway,left] {$\W $};
\node[rectangle, draw=black,fill=green!10!,minimum size=10,line width=1,text width=15]at(1,2){$+\frac{\pi}{2}$};
\draw[thick,->,>=latex](6,-1)--(7,-1)node[below,midway]{Z-axis};
\draw[thick,->,>=latex](6,-1)--(6,0)node[above,midway,rotate=90]{Y-axis};
\end{tikzpicture}
\caption{Schematics of shot noise limited interferometer. LS: Laser source, D1,D2: Photo detectors. The yellow rectangle represents three level atomic medium which is placed in a running wave optical cavity formed by mirrors 1,2 and by a semi-transparent mirror `m'. The atomic level structure and the interaction of laser fields with the atoms in the atomic medium is shown in the big yellow circle. The red circle represents the cross-section area of classical driving field which is coupling $|a\ket-|c\ket$ transition and propagating perpendicular to YZ-plane. The field inside the running-wave cavity, represented by $\hat c$, is coupling $|a\ket-|b\ket$ transition.}\label{f1}
\end{figure*}
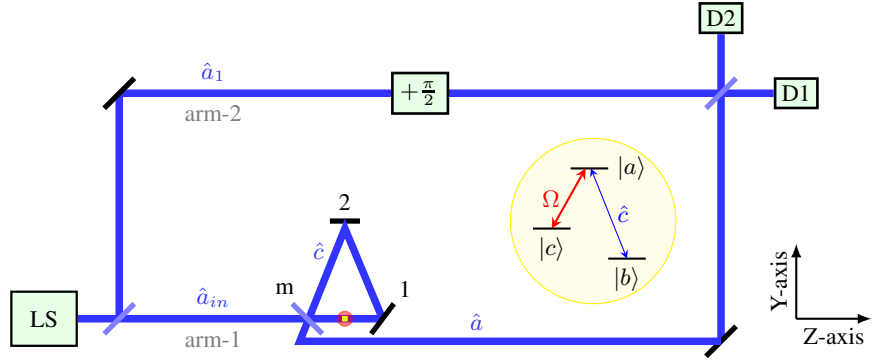

Figure-\ref{f1} shows the schematics of the back-action evasion interferometer.
A three-level atomic medium (shown by small yellow rectangle in \fig{f1}), with atomic levels $|a\ket,$ $|b\ket$, and $|c\ket,$ is placed inside a running wave optical cavity formed by a semi-transparent mirror `m' and two mirrors 1\& 2 as shown in \fig{f1}. The running wave optical cavity, along with atomic medium, is placed in one of the arms of the Mach-Zehnder interferometer as shown in \fig{f1}. The atomic medium is enclosed in a small volume such that no atoms enter or leave the interaction region. The energy level structure of the atoms inside the atomic medium is shown in the big yellow circle in \fig{f1}. The cavity field (shown in blue color) couples $|a\ket-|b\ket$ transition while a strong classical field, whose area of cross-section is shown by red circle, propagating perpendicular to the YZ-plane couple $|a\ket-|c\ket$ transition.

The Hamiltonian\cite{lukin,vernac,yavuz,dantan-04} $\hat H$ is written as
\ba\bs\hat H=\um{j}{N}\frac{(\hat p^j)^2}{2m}+\hbar\nu(\td{\hat{c}}^\+\td{\hat{c}}+\frac{1}{2})+\um{j}{N}\um{u}{a,b,c}\hbar\w_u\hat\s_u^j\\+\big(\um{j}{N}\hbar|a\ket^j\bra c|^j\td\Omega+\um{j}{N}\hbar g|a\ket^j\bra b|^j(\hat z^j)\td{\hat{c}}+\text{H.C}\big)\\+i\hbar\sqrt\z(\td{\hat c}^\+\td{\hat a}_{in}-\td{\hat a}_{in}^\+\td{\hat c})+\um{j}{N}F\cdot\hat z^j,\label{1}
\end{split}\ea
where H.C is the hermitian conjugate, the superscript $j$ indicates $j$-$th$ atom and $N$ is the number of atoms. $\hat p^j$ is the momentum of the `$j$' atom along Z-axis and $m$ is atom's mass. $\td{\hat c}$ and $\nu$ are the annihilation operator and frequency of the cavity field. $\hbar$ is reduced Planck constant, $\w_u$ is the eigen frequency of the atomic level $|u\ket$ when the atoms are at rest and $\hat{\s}_u^j=|u\ket^j\bra u|^j$. $g$ is the coupling constant between weak cavity field $\hat c$ and $|a\ket-|b\ket$ transition. $\td\W$ is the Rabi frequency of the strong classical driving laser propagating along X-axis. Note that $|a\ket^j\bra c|^j$ has no $\hat z$-dependence in \eq{1} as the driving field perpendicular to Z-axis. $\hat a_{in}$ and $\zeta/2$ are the input and decay rate of semi-transparent mirror `m', respectively. $F$ is the classical force acting on the atom, and $\hat z^j$ is the displacement of `$j$' atom.

We separate the $\hat z^j$ dependence from $|a\ket^j\bra b|^j(\hat z^j)$ by writing $|a\ket^j\bra b|^j(\hat z^j)=|a\ket^j\bra b|^je^{ik\hat z^j}$, where $k$ is the wave-vector of $\hat c$. Equations of motion for atom-cavity field interaction, after rotating wave approximation, are given as
\bsa\ba\dot{\hat c}=-\frac{\z}{2}\hat c-i\um{j}{N}g^*\hat\s_{ba}^j(\hat z^j)+\sqrt\z\hat a_{in}\ea
\ba\dot{\hat\s}_{ba}^j(\hat z^j)=\G^j\hat\s_{ba}^j(\hat z^j)+ig(\hat\s_{a}^j-\hat\s_{b}^j)\hat c-i\W\hat\s_{bc}^j(\hat z^j)+\hat F_{ba}^j,\ea
\ba\dot {\hat\s}_{bc}^j(\hat z^j)=\G_o^j\hat\s_{bc}^j(\hat z^j)+ig\hat\s_{ac}^j\hat c-i\W^*\hat\s_{ba}^j(\hat z^j)+\hat F_{bc}^j,\ea
\ba\dot{\hat\s}_{ac}^j=-\g\hat\s_{ac}^j+i(\hat\s_{c}^j-\hat\s_{a}^j)\W^*+ig^*\hat\s_{bc}^j(\hat z^j)\hat c^{\+}+\hat F_{ac}^j,\ea
\ba\dot{\hat p}^j=i\hbar k(g^*\hat c^\+\hat\s_{ba}^j(\hat z^j)-g\hat\s_{ab}^j(\hat z^j)\hat c)+F,\el{2e}\esl{2}
where $\G^j=i(\frac{k\hat p^j}{m}-\frac{\hbar k^2}{2m})-\g$, $\G_o^j=\frac{k\hat p^j}{m}-\frac{\hbar k^2}{2m}-\g_o,$ $|c\ket^j\bra a|^j=\hat\s_{ca}^je^{-i\nu_{d}t}$ ($\nu_d$ is the driving laser frequency), $|b\ket^j\bra a|^j(\hat z^j)=\hat\s_{ba}^j(\hat z^j)e^{-i\nu t},$ $|b\ket^j\bra c|^j(\hat z^j)=\hat\s_{bc}^j(\hat z^j)e^{-i(\nu-\nu_d)t},$ $\td\W=\W e^{-i\nu_dt}$, $\td{\hat{c}}=\hat ce^{-i\nu t}$. We assumed that the cavity field and the driving field are on resonance with $|a\ket-|b\ket$ and $|a\ket-|c\ket$ transitions, respectively, when the atom is at rest. $\hat F^j_{uv}$ is the slowly varying noise operator, $\g$ is the decoherence on transitions $|a\ket-|b\ket$ and $|a\ket-|c\ket$. $\hat a_{in}$ is the slowly varying operator of $\td{\hat a}_{in}.$ On the electric dipole forbidden transition $|b\ket-|c\ket$, decoherence $\gamma_o$ is added phenomenologically. The first term on the right hand side of \eq{2e} represents the radiation pressure force~\cite{wineland} because of absorption of cavity field. From now on, we drop the recoil frequency $\hbar k^2/2m$ term in \eq{2} by assuming that it is much smaller than the Doppler shift term $\hat p^jk/m$. 

Assuming that cavity field as weak field, we treat $\hat c$ up to its first order while keeping the $\W$ to all orders (the superscript `0' indicates zeroth order in $\hat c$, while the superscript `1' indicates first order in $\hat c$). Now the relevant equations of motion are given as
\bsa\ba\hat a=\hat a_{in}-\sqrt\z\hat c,\el{3a}
\ba\dot{\hat c}=-\frac{\z}{2}\hat c-i\um{j}{N}g^*\hat\s_{ba}^{j(1)}(\hat z^{j(0)})+\sqrt\z\hat a_{in}\el{3b}
\ba\dot{\hat\s}_{ba}^{j(1)}(\hat z^{j(0)})=\G^{j(0)}\hat\s_{ba}^{j(1)}(\hat z^{j(0)})-ig\hat c-i\W\hat\s_{bc}^{j(1)}(\hat z^{j(0)})+\hat F_{ba},\el{3c}
\ba\dot {\hat\s}_{bc}^{j(1)}(\hat z^{j(0)})=\G^{j(0)}_o\hat\s_{bc}^{j(1)}(\hat z^{j(0)})-i\W^*\hat\s_{ba}^{j(1)}(\hat z^{j(0)})+\hat F_{bc},\el{3d}
\ba\dot{\hat p}^{j(0)}=F,\el{3e}\esl{3}
where $\G^{j(0)}=ik\hat p^{j(0)}/m-\g,$ $\G^{j(0)}_o=ik\hat p^{j(0)}/m-\g_o,$ $\hat a$ is the output field~\cite{gardiner,collet} from `m'. In writing \eq{2}, I used the EIT system properties that all the atomic population resides in $|b\ket$, and hence $\hat\s^{j(0)}_{b}=1$ while $\hat\s^{j(0)}_{a}=\hat\s^{j(0)}_{c}=\hat\s^{j(0)}_{ac}=\hat{\s}_{ba}^{j(0)}=\hat{\s}_{bc}^{j(0)}=0$\cite{dantan-05,amy,junxiang,sun,sankar-ps}. Equation~(\ref{3e}) implies that the time evolution of $\hat p^{j(0)}$ is not disturbed by the measuring device and hence momentum $\hat p^{j(0)}$ is a quantum non-demolition variable. Its worth noting that back-action is present in \eq{2e} and it is the EIT system which leads to back-action evasion in \eq{3}. Application of EIT for velocity read out purpose has been studied~\cite{sankar-pra1, kuan, rabl,safari}, however, the possibility of back-action evasion is not shown explicitly. By solving \eq{3e}, we can replace $k\hat p^{j(0)}/m$ and $\hat z^{j(0)}$ with classical values $Fkt_m/m$ and $z^{j(0)}$, respectively, where $t_m$ is the time of measurement. Throughout this manuscript, we assume that $t_m$ is much less than the characteristic time an atom takes to travel from one end of the atomic medium to the other, along Z-axis.
\subsection{Signal}
Assuming that $t_m\ll 1/\g$, mean value of $\hat c$ can be obtained by solving \eq{3} as
\ba\dot{\bar c}=
-\frac{\sqrt{\z}\bar a_{in}}{\A_o-\frac{\z}{2}},\el{8}
where  \ba\bs\A_o=\frac{N|g|^2(i\frac{kt_m}{m}F-\g_o)}{(i\frac{kt_m}{m}F-\g)(i\frac{kt_m}{m}F-\g_o)+|\W|^2},\end{split}\el{7}
$\bar c$ and $\bar a_{in1}$ are classical mean values of $\hat c$ and $\hat a_{in}$, respectively. Assuming that $\W^2\gg\g\g_o$, and by considering $\A_o$ only up-to the first order of $Fkt_m/m$, we can approximate
the output from semitransparent mirror `m' as
\ba\bar a
\approx\frac{(iF\frac{kt_m}{m}\frac{N|g|^2}{|\W|^2}\z+\frac{N^2|g|^4\g_o^2}{|\W|^4}-\frac{\z^2}{4})}{(\frac{N|g|^2\g_o}{|\W|^2}+\frac{\z}{2})^2}\bar a_{in},\el{9}
where $\bar a$ is the mean value of $\hat a$. Electromagnetic field in the arm-2 of \fig{f1} is represented by $\hat a_1$. Hence the difference in the photo detector readings, after adding a constant $\pi/2$ phase\cite{caves} to $\hat a_1$, is given as
\ba\hat I_1-\hat I_2=\hat a^\+\hat a_1+\hat a\hat a_1^\+,\el{10}
where $\hat I_1$ and $\hat I_2$ are the intensities at photo detectors D1 and D2, respectively.
 By using \eq{9} and the relation $\bar a_1=i\bar a_{in}$, we can write mean value of \eq{10} as
\ba\bra I_1-I_2\ket=
\frac{-2\frac{kt_m}{m}F\z|\bar a_{in}|^2}{(N|g|^2\g_o/|\W|^2+\frac{\z}{2})^2}\frac{N|g|^2}{|\W|^2}\el{12}
Equation-\ref{10} has maximum value when $N|g|^2\g_o/|\W|^2=\z/2$. Hence the maximum signal we can obtain is
\ba\bra I_1-I_2\ket=\frac{-kt_mF}{m\z}\frac{\zeta}{\gamma_o}|\bar a_{in}|^2.\el{13}
\subsection{Noise spectrum}
The linearized equations of motion for the fluctuations are
\begin{subequations}\ba
\frac{\dho}{\dho t}\hat\de_c=-\frac{\z}{2}\hat\de_c-i\um{j}{N}g^*\hat\de^{j(1)}_{ba}+\sqrt\z\hat\de_{in},\el{14a}
\ba\dot{\hat{\de}}^{j(1)}_{ba}=(i\frac{kt_m}{m}F-\g)\hat\de^{j(1)}_{ba}-ig_c\hat c-i\W\hat\de^{j(1)}_{bc}+\hat {F}_{ba}^j,\ea
\ba\dot{\hat\de}^{j(1)}_{bc}=(i\frac{kt_m}{m}F-\g_o)\hat\de^{j(1)}_{bc}-i\W^*\hat\de^{j(1)}_{ba}+\hat{F}_{bc}^j,\el{14}\end{subequations}
where $\hat \de_c$, $\hat\de_{ba}^{j(1)}$, $\hat\de_{bc}^{j(1)}$, $\hat\de_{in}$ are the quantum fluctuation of $\hat c$, $\hat\s_{ba}^{j(1)}(\hat z^{j(0)})$, $\hat\s_{bc}^{j(1)}(\hat z^{j(0)})$, and $\hat a_{in}$, respectively. The terms $kt_mF/m,$ $\hat\de^{j(1)}_{bc},$ and $\hat\de^{j(1)}_{ab}$ are very small, hence the product terms such as $ikt_mF\hat\de^{j(1)}_{bc}/m$, can be neglected in \eq{14}. By using the Fourier transform function $\mathfrak{F}(x(t))=\frac{1}{\sqrt{2\pi}}\inti x(t)e^{i\w t}\dv t$, the solution to the above equations in the Fourier frequency space is given as
\ba\hat\de_c(\w)=-\frac{\hat F(\w)+\sqrt\z\hat \de_{in}(\w)}{i\w-\frac{\z}{2}+\frac{N|g|^2(i\w-\g_o)}{(i\w-\g)(i\w-\g_o)+|\W|^2}},\el{18}
where $$\hat F(\w)=\um{j}{N}g^*\frac{-\W\hat F_{bc}^j(\w)+i(i\w-\g_o)\hat F_{ba}^j(\w)}{(i\w-\g)(i\w-\g_o)+|\W|^2}.$$ Fluctuation in output field is given as
\ba\bs\hat\de(\w)=
\frac{\hat\de_{in}(\w)(i\w+\A(\w)+\frac{\zeta}{2})+\sqrt\z\hat F(\w)}{i\w+\A(\w)-\frac{\z}{2}},\end{split}\el{19}
where $\hat\de$ is the fluctuation in $\hat a$, and $$\A(\w)=\frac{N|g|^2}{(i\w-\g+\frac{|\W|^2}{i\w-\g_o})}.$$ Hence, we can write
\ba\bra\hat{\de}(\w)\hat{\de}^\+(\w)\ket=\frac{|i\w+\A(\w)+\frac{\zeta}{2}|^2\bra\hat\de_{in}(\w)\hat\de_{in}^\+(\w)\ket+\z\bra \hat F(\w)\hat F^\+(\w)\ket}{|i\w+\A(\w)-\frac{\z}{2}|^2}.\el{21}
Fluctuation in \eq{10}, denoted by $\hat\De$, is given as
\ba\hat\De=\hat I_1-\hat I_2-\bra\hat I_1-\hat I_2\ket=\hat{\de}^\+_1\bar a+\hat\de_1\bar a^*+\hat\de^\+\bar a_1+\hat\de\bar a_1^*.\el{22}
Variance of $\hat\De(\w),$ where $\hat\De(\w)=\mathfrak{F}(\hat\De)$, gives the power spectral density $V^2$ as 
\ba\bs\bra \hat\De^\+(\w)\hat\De(\w')\ket=|\bar a|^2\bra \hat{\de}_1(\w)\hat{\de}_1^\+(\w')\ket+|\bar a_1|^2\bra \hat{\de}(\w)\hat{\de}^\+(\w')\ket\\
=V^2\de(\w+\w'),\end{split}\el{23}
where $V^2$ can be computed by using the correlation functions
\bsa\ba\um{j,j'}{N}\bra \hat F_{bc}^j(t_1)\hat F_{bc}^{j'\+}(t_2) \ket
=2N\gamma_o\de(t_1-t_2),\ea
\ba\um{j,j'}{N}\bra \hat F_{ba}^j(t_1)\hat F_{ba}^{j'\+}(t_2) \ket
=2N\gamma\de(t_1-t_2),\ea
\ba\bra \hat\de_{in}(t_1)\hat\de_{in}^\+(t_1)\ket
=\de(t_1-t_2).\el{20}\esa
At $\w=0$, by substituting \eq{21}, \eq{8}, and $\bar a_1=i\bar a_{in}$ in \eq{23}, we evaluate
\begin{widetext}
\ba V=\sqrt{\Big(\frac{((\frac{N|g|^2\g_o}{|\W|^2})^2-(\frac{\z}{2})^2)^2}{|\frac{N|g|^2\g_o}{|\W|^2}+\frac{\z}{2}|^4}+\frac{|\frac{-N|g|^2\g_o}{|\W|^2}+\frac{\z}{2}|^2+\z(2\frac{N|g|^2\g_o}{|\W|^2})}{|-\frac{N|g|^2\g_o}{|\W|^2}-\frac{\z}{2}|^2}\Big)|\bar a_{in}|^2}.\el{24}\end{widetext}
At $N|g|^2\g_o/|\W|^2=\z/2$, we have
\ba V=\sqrt{|\bar a_{in}|^2}.\el{25}
\section{Thermal noise}
Until now, we have neglected the intrinsic motion of atoms in a gas. At temperature $T$, the atoms in the gas move randomly in all the directions because of thermal energy. The random thermal motion leads to Doppler broadening~\cite{wineland-79} which can be estimated by using Maxwell-Boltzmann distribution. Since the probe and drive lasers are in $z$ and $x$ directions, respectively, we only need to consider the thermal Doppler detuning because of $v_z$ and $v_x$. Where $v_z$ and $v_x$ are the thermal velocities of the atom with respect to probe and drive laser fields, respectively. So the effect of temperature can be accounted by using the Maxwellian velocity distribution as
\begin{widetext}
\ba\frac{m}{2\pi k_BT}\inti\inti e^{-m(v_z^2+v_x^2)/2k_BT}\um{j}{}\hat\s_{ba}^{j(1)}(v_x,v_z)\dv v_z\dv v_x=\frac{m}{2\pi k_BT}\inti\inti e^{-m(v_z^2+v_x^2)/2k_BT}\Big(N\bar\s_{baD}^{(1)}+\mathfrak{F}^{-1}\Big(\um{j}{}\hat\de^{j(1)}_{baD}(\w)\Big)\Big)\dv v_z\dv v_x,\el{26}
where
\bsa\ba\bar\s^{(1)}_{baD}=\frac{ig\bar a(iF\frac{kt_m}{m}+i(\w_{ab} v_z-\w_{ac}v_x)/c-\g_o)}{(iF\frac{kt_m}{m}+i\w_{ab}v_z/c-\g)(iF\frac{kt_m}{m}+i(\w_{ab} v_z-\w_{ac}v_x)/c-\g_o)+|\W|^2},\ea
\ba\hat\de^{j(1)}_{baD}(\w)=\frac{iNg\hat{\de}_c(\w)}{(i\w+i\frac{\w_{ab}v_z}{c}-\g)+\frac{|\W|^2}{i\w+i\frac{(\w_{ab}v_z-\w_{ac}v_x)}{c}-\g_o}}-\frac{\um{j}{N}\Big(\frac{i\W\hat{F}_{bc}^j(z,\w)}{(i\w+i\frac{\w_{ab}v_z-\w_{ac}v_x}{c}-\g_o)}+\hat{F}_{ba}^j(z,\w)\Big)}{(i\w+i\frac{(\w_{ab}v_z-\w_{ac}v_x)}{c}-\g)+\frac{|\W|^2}{i\w+i\frac{(\w_{ab}v_z-\w_{ac}v_x)}{c}-\g_o}},\ea\esl{27}\end{widetext}
where $k_B$ is Boltzmann constant, $\w_{ab}=\w_a-\w_b$ and $\w_{ac}=\w_a-\w_c.$ It should be noted that \eq{27} can be approximated as a linear function of $v_x$ and $v_z$ when 
\ba\bs|\W|^2&\gg i\w_{ab}v_z\g_o/c+i(\w_{ab}v_z-\w_{ac}v_x)\g/c\\&\gg\w_{ab}v_z(\w_{ab}v_z-\w_{ac}v_x)/c^2.\end{split}\el{28} 

The signal and noise given in \eq{13} and \eq{25}, respectively, are evaluated when $|\W|^2=2N|g|^2\g_o/\z$. Hence
\ba|\W|^2=2N|g|^2\frac{\g_o}{\z}
=6\pi\frac{N}{V}\frac{c^3\g}{\w_{ab}^2}\frac{\g_o}{\z},\el{29}
where $N/V$ represents the density of atoms in the running wave cavity.

By considering realistic parameters: $N/V\approx10^{18}$\,m$^{-3}$, $c=3\times10^8$\,m/s, $\g=10^7$\,Hz, $\w_{ab}=5\times 10^{15}$\,Hz, $\g_o=10^3$\,Hz, $\z=10^6$\,Hz, we estimate that $|\W|^2\approx3\times 10^{16}$\,Hz$^2.$  By considering that $\w_{cb}=10^{-6}\w_{ab},$ $m\approx 1.4\times 10^{-25}$\,Kg, $k_B=1.3\times 10^{-23}$\,J/K, and $T=1$\,K, the Doppler width of $|a\ket-|b\ket$ transition is $\w_{ab}\sqrt{2\ln2k_BT/mc^2}\approx2.8\times10^8$\,Hz and the Doppler width of $|c\ket-|b\ket$ transition is $\w_{cb}\sqrt{2\ln2k_BT/mc^2}\approx264$\,Hz. Hence \eq{28} is practically fulfilled~\cite{fleischhauer-94} when the temperature of the atomic medium is sufficiently low and then we can approximate \eq{27} as   
\begin{widetext}
\bsa\ba\bar\s^{(1)}_{baD}= ig\bar a\Big(\frac{(i[\w_{ab}\frac{v_z}{c}-\w_{ac}\frac{v_x}{c}]-\g_o)}{|\W|^2}+i\frac{F\frac{kt_m}{m}}{|\W|^2}\Big).\ea
\ba\hat\de^{j(1)}_{baD}=\frac{iNg\hat{\de}_c(\w)(i\w+i\frac{(\w_{ab}v_z-\w_{ac}v_x)}{c}-\g_o)}{(i\w-\g)(i\w-\g_o)+|\W|^2}-\frac{\um{j}{N}\Big(i\W\hat{F}_{bc}^j(z,\w)+(i\w+i\frac{\w_{ab}v_z-\w_{ac}v_x}{c}-\g_o)\hat{F}_{ba}^j(z,\w)\Big)}{(i\w-\g)(i\w-\g_o)+|\W|^2}.\ea\esl{30}
By substituting \eq{30} in \eq{26}, we will see that
\ba\frac{m}{2\pi k_BT}\inti\inti e^{-m(v_z^2+v_x^2)/2k_BT}\um{j}{}\hat\s_{ba}^{j(1)}(v_x,v_z)\dv v_z\dv v_x=\um{j}{N}\hat\s_{ba}^{j(1)}.\el{31}\end{widetext}
Hence the Doppler broadening or the thermal noise effect is practically zero when the condition given in \eq{28} is fulfilled.
\section{Discussion}
By comparing the noise in \eq{25} with signal in \eq{13}, we estimate the force sensitivity $F_s$ as
\ba F_s=\frac{V}{\bra \hat I_1-\hat I_2\ket}F=\frac{m\g_o}{kt_m}\frac{1}{\sqrt{|\bar a_{in}|^2}}\el{32}
Recently, optomechanical systems~\cite{meystre} has gained a lot of attention as ultra-sensitive force detectors~\cite{li,sankar-pra3,korppi,sumei-fp}. Generally, classical force is estimated in optomechanical systems by measuring the position~\cite{wilson} of the object on which force is acting. However, accuracy of position measurement intrinsically leads to quantum back-action noise~\cite{caves-rp} which limits the best precision achievable. Application of squeezed states~\cite{woolley,xu} and back-action evasion~\cite{korppi,bohnet,lecocq,motazedifard,tsang} methods are proposed to over come SQL, but still, thermal noise~\cite{sankar-njp1,anetsberger,wilson,teufel} limits the sensitivity of optomechanical systems. Moreover, preparing optimum squeezed states to overcome the SQL is experimentally challenging. In this manuscript we proposed an interferometry technique in which both the back-action noise and the thermal noise are suppressed. 
\section{Conclusion}
A technique to measure classical force without measurement back-action noise and thermal noise is described. Back-action evasion is achieved by using EIT phenomena.  Role of temperature is studied and derived the sufficient conditions for nullifying the thermal noise.  With no thermal noise and no back-action noise, shot noise is the only limitation in this measurement scheme.
\section{Acknowledgements}
S.D. thanks Prof. Peng Zhang for fruitful discussions. This work is supported by the Science Challenge Project (under Grant No. TZ2018003) and the Natural Science Foundation of China (under Grants No. 11774024, No. 11534002, and No. U1530401).
\bibliography{references}
\end{document}